\RequirePackage{fix-cm}

\documentclass[natbib,smallextended]{svjour3}

\usepackage{graphicx}
\usepackage{hyperref}
\usepackage{url}
\usepackage{comment}
\usepackage{todonotes}
\usepackage{color}
\usepackage{multirow}
\usepackage[most]{tcolorbox}
\usepackage{tabularx}
\usepackage{balance}
\usepackage{changepage}

\begin{document}

\title{Testing Research Software: A Survey}

\author{Nasir U. Eisty         \and
        Jeffrey C. Carver 
}

\institute{N. U. Eisty \at
            Department of Computer Science\\
            Boise State University \\
            Boise, ID, USA \\
            \email{nasireisty@boisestate.edu}
           \and
           J. C. Carver \at
              Department of Computer Science\\
              University of Alabama\\
             Tuscaloosa, AL, USA\\
              \email{carver@cs.ua.edu}
}

\date{Received: Sept. 28, 2021 / Accepted: May 26, 2022}

\maketitle

\begin{abstract}
\textit{Background:}
Research software plays an important role in solving real-life problems, empowering scientific innovations, and handling emergency situations.
Therefore, the correctness and trustworthiness of research software are of absolute importance.
Software testing is an important activity for identifying problematic code and helping to produce high-quality software.
However, testing of research software is difficult due to the complexity of the underlying science, relatively unknown results from scientific algorithms, and the culture of the research software community.
\textit{Aims:}
The goal of this paper is to \textit{better understand current testing practices, identify challenges, and provide recommendations on how to improve the testing process for research software development.} 
\textit{Method:}
We surveyed members of the research software developer community to collect information regarding their knowledge about and use of software testing in their projects.
\textit{Results:}
We analysed 120 responses and identified that even though research software developers report they have an average level of knowledge about software testing, they still find it difficult due to the numerous challenges involved. 
However, there are a number of ways, such as proper training, that can improve the testing process for research software.
\textit{Conclusions:}
Testing can be challenging for any type of software.
This difficulty is especially present in the development of research software, where software engineering activities are typically given less attention.
To produce trustworthy results from research software, there is a need for a culture change so that testing is valued and teams devote appropriate effort to writing and executing tests.  

\keywords{Software Testing \and Survey \and Research Software \and Software Engineering}

\end{abstract}

\section{Introduction}
\label{sec:introduction}
Research software is software developed by researchers from a wide variety of domains including, but not limited to, science, engineering, business, and humanities~\citep{8588655}.
Research software can serve different purposes.
Sometimes researchers develop software to make predictions about or to better understand real-world processes~\citep{Kanewala:2014:TSS:2658281.2658307}.
This type of research software solves computationally complex or data-intensive problems.
For example, researchers develop software for large-scale physical phenomena, such as weapons or medical simulations, that run on high-performance computers.
Researchers also develop smaller simulations that run on desktop machines or small cluster~\citep{Kelly08thechallenge}. 
Research software can provide infrastructure support (e.g., messaging middleware or scheduling software) or libraries for mathematical and scientific programming (e.g., linear algebra or symbolic computing).

The last few years have seen significant growth in research software development.
While researchers have been building software to support their research for many decades, their primary goals relate to the research outputs, not to the quality or sustainability of the underlying software~\citep{5337646}.
More recently, the research software community has begun to recognize the need for focusing on the quality and sustainability of their software~\citep{8565942}.
Increasingly, the development of research software requires the use of advanced skill-sets that developers must build over time~\citep{2255589632}. 
These developers need to use best practices to ensure the reliability and sustainability of the software they develop.
Therefore, it is the right time to investigate how research software developers use a key practice like testing to identify the challenges and develop best practices for research software.

The quality of research software is critical.
Researchers use research software in mission-critical situations and decision making~\citep{Kanewala:2014:TSS:2658281.2658307}. 
In addition, researchers use the results from research software as evidence in research publications~\citep{4548404,doi:10.1177/1094342005056094}.
Therefore, it is important that research software have a correct design and implementation. 
Low quality software produces less trustworthy results and is prone to failure in mission-critical situations.
Software quality problems have even caused scientists to retract publications~\citep{Miller1856}.
Therefore, researchers need to develop high-quality software that produces trustworthy results and functions properly in critical situations.

The term \textit{research software developer} refers to a person who develops research software.
Research software is most often produced by researchers themselves, often within academia, by faculty, staff, postdocs, and students.
Research software developers range from researchers who possess little or no software engineering knowledge to experienced professional software developers with considerable software engineering knowledge~\citep{Kanewala:2014:TSS:2658281.2658307}.

Testing is a useful practice for producing high-quality software.
Unfortunately, because of its complex computational behavior, it is very difficult to test research software.
Research software developers often build software based upon a set of mathematical equations and use mathematical analysis to verify the corrections of the computational model~\citep{5228715,doi:10.1177/1094342004048534}.
For example, researchers use research software to determine the impact of modifications to nuclear weapon simulations since real-world testing is too dangerous and not allowed~\citep{doi:10.1177/1094342004048534}.

While testing is a useful practice, there are some technical challenges for testing research software.
The first challenge is the lack of test oracles~\citep{5069163}.
An oracle is pragmatically unattainable in most of the cases for research software because researchers develop software to find previously unknown answers.
Due to the lack of test oracles, research software developers often use judgment and experience to check the correctness of the software. 
The second challenge is the large number of tests required to test research software using standard testing techniques.
Also, the large number of input parameters makes it challenging to manually selecting a sufficient test suite~\citep{10.1007/978-3-540-69389-5_34}.
Finally, the presence of legacy code makes testing research software very challenging~\citep{5999647}.

Because of these challenges, research software developers are unlikely to use systematic testing to check the correctness of their code~\citep{5069163,article2580,Kelly08thechallenge}.
Even though these developers conduct validation checks to ensure the software correctly models the physical phenomenon of interest~\citep{5228715,article2580},
there is still a need for testing that identifies differences between the model and the code~\citep{gmd-4-435-2011}.
In addition, sometimes the reason for limited use of systematic testing results from the testing challenges posed by the software itself~\citep{Easterbrook:2010:CCG:1882362.1882383}. 

Because of the various challenges to properly and fully testing research software, there is a need to better understand how research software developers actually perform testing activities in practice.
Therefore, the goal of this paper is to \textit{better understand the testing process, identify challenges, and provide recommendations on how to improve the testing process in research software development.}
To make this high-level goal more tractable,  Section~\ref{sec:research_questions} reviews related work and poses a set of specific research questions.
Then, to answer these questions, the paper describes the results from our survey of research software developers.

Based on the results from the 120 respondents to the survey, the key contribution of this paper are:
\begin{itemize}
    \item An overview of the level of knowledge research software developers have on software testing;
    \item A description of the current testing practices used in the research software community;
    \item A list of the difficulties of testing research software;
    \item An analysis of the compatibility of commercial/IT testing techniques to research software; and
    \item An identification of areas of improvement in the testing process for research software.
\end{itemize}

\section{Research Questions}
\label{sec:research_questions}
This section discusses the research questions with related work.
We use the related work to motivate a series of research questions that will ultimately drive the survey design.

\vspace{4pt}
\noindent
\textbf{RQ1: What level of knowledge do research software developers have about software testing?}

Research software developers often have little or no formal software engineering knowledge~\citep{Easterbrook:2010:CCG:1882362.1882383,5337646,5069155}.
There is a lack of recognition for the skills and knowledge required for software development~\citep{7739651}.
These developers are typically unfamiliar with available testing methods~\citep{5337644,5069155}.
As a consequence, they do not usually have a set of written quality goals.
Researchers even treat software development as a secondary activity.
Therefore, we pose this research question to better understand the level of knowledge research software developers possess about software testing and verify the claims from the literature.

\vspace{4pt}
\noindent
\textbf{RQ2: How do research software developers test software?}

The literature only provides a few examples of how research software developers perform testing.
Research software developers commonly omit unit testing because they have misconceptions about the benefits and difficulties of implementing unit tests~\citep{5999647}.
Research software developers under-utilize verification testing because they are unaware of the need for it and the methods for applying it~\citep{5337646}.
Research software projects often do not even include automated acceptance testing and regression testing~\citep{Nguyen-Hoan:2010:SSS:1852786.1852802}.
Furthermore, the research software community is lagging in the use of available testing tools, at least partially due to the wide use of FORTRAN~\citep{4548404,5337646,6258305}.
Therefore, to better understand how research software developers actually perform testing, we pose this research question.

\vspace{4pt}
\noindent
\textbf{RQ3: Why is testing research software difficult?}

Testing research software is challenging.
A previous systematic literature review (SLR) reported testing challenges due to the characteristics of research software and the cultural differences between researchers and more traditional software engineers~\citep{Kanewala:2014:TSS:2658281.2658307}.
The authors subdivided the testing challenges resulting from the characteristics of research software into four categories: a) test case development, b) producing expected test case output values, c) test execution, and d) test results interpretation.
Then subdivided the testing challenges resulting from the cultural differences between research software developers and more traditional software engineers into three categories: a) limited understanding of testing concepts, b) limited understanding of the testing process, and c) not applying known testing methods.
Because the SLR was only able to capture and report on challenges actually published in the literature, to better evaluate whether these challenges, and others, happen in practice, we pose this research question.

\vspace{4pt}
\noindent
\textbf{RQ4: Is it possible to adapt existing testing methods to test research software?}

Based on the SLR described in the previous research question~\citep{Kanewala:2014:TSS:2658281.2658307}, there is little evidence that research software projects use the available testing methods as we describe below. 
Only a few projects mentioned the use of unit testing~\citep{doi:10.1177/1094342005056094,gmd-4-435-2011,4548407,5467013}.
There is only one study that mentioned the use of integration testing~\citep{doi:10.1177/1094342005056094}.
A few other studies mentioned the use of system testing~\citep{gmd-4-435-2011,5999647}, acceptance testing~\citep{5999647}, and regression testing~\citep{doi:10.1177/1094342005056094,gmd-4-435-2011,4476223}.
There are a lot of research software projects in existence, but the number that appear in literature about testing is very low~\citep{Kanewala:2014:TSS:2658281.2658307}.
Therefore, to understand whether existing testing methods could be adapted to address this lack of software testing in research software, we pose this research question.

\vspace{4pt}
\noindent
\textbf{RQ5: What improvements to the testing process do research software developers need?}

To better identify ways to advance the testing of research software, there is a need to better understand specific ways to improve the testing process and overcome the testing challenges that exist.
Because of the challenges, developers often do not want to write tests.
Therefore, addressing the testing challenges can make the process welcoming to research software developer.
For example, developing a test oracle is a key challenge for testing research software.
Some research software projects have addressed this challenge through creating pseudo oracles, which is code developed separately to produce the intended output given the same input as the original program~\citep{gmd-4-435-2011,5337646,Nguyen-Hoan:2010:SSS:1852786.1852802}. 
Though creating a pseudo oracle is not commonly applied to all research software projects but this technique may work for some.
Another challenge for testing research software is obtaining adequate budget~\citep{Nguyen-Hoan:2010:SSS:1852786.1852802,7739651,Segal2009}.
Projects may be able to overcome this challenge through increasing awareness among project decision-makers of the need for providing a budget for testing activities.
To identify other approaches for improving the current testing process in research software development, we pose this last research question.

\section{Methodology}
\label{sec:methodology}
To answer the research questions described in Section~\ref{sec:research_questions}, we conducted an online questionnaire survey of members of the research software development community.
 
To reach a broad audience and produce useful results, we decided to conduct a survey because surveys are useful for describing the characteristics of a large population.
Section~\ref{sec:Survey-Design} describes the design of the study.
Section~\ref{sec:Data-Analysis} explains the process we followed to analyze the results of the survey.

\subsection{Survey Design}
\label{sec:Survey-Design}

\subsubsection{Questionnaire}
Using the research questions, we enumerated the survey questions shown in Figures~\ref{fig_survey_questions1} and~\ref{fig_survey_questions2}. 
Note the bold headings that enumerate the Research Questions did not appear in the survey, but are included here for clarity. 
Because we anticipated survey respondents may not be familiar with standard software engineering definitions, we defined key software engineering concepts on the survey.
We provided definitions for the roles listed in Q2, the testing methods listed in Q11, and the techniques listed in Q13 in the survey (see Appendix~\ref{sec:Appendix}) to address the possibility that a respondent may be familiar with a concept but unaware of the proper term.
In addition, to ensure respondent had the proper context we included the following statement at the beginning of the survey: ``When answering these questions, please consider the research software project on which you are the most active.''

\subsubsection{Pilot Study}
We piloted the initial survey with three experienced research software developers from Los Alamos National Laboratory (LANL), where the first author was an intern.
The pilot testers each work in a different scientific domain: Computational Physics, Computational Material Science, and Applied Mathematics.
Each pilot tester has more than 15 years of experience developing and using software for their research. 
Based on their feedback, we reworded some questions for clarity to research software developers. 
We then deployed the survey via the Qualtrics platform.

\subsubsection{Distribution}
We used the following solicitation methods to reach a broad population of research software developers. 
First, we sent the survey to mailing lists that reach research software developers.
Those lists include the prior attendees of the SE4Science workshops\footnote{https://se4science.org/workshops/} along with a custom-made email list of contributors to research software repositories mined from GitHub.
Second, we advertised the survey in two Slack channels, one for international research software engineers (\textit{RSE}) and one focused on research software engineers in the US (\textit{US-RSE}).
The subscribers to  these slack channels are research software engineers that range from graduate students to experienced researchers working in academia or research labs.
Third, we advertised the survey in the monthly newsletter of Better Scientific Software (BSSw) and the IDEAS-productivity mailing list.
The participants in these mailing lists are research scientists, faculty members, graduate students, and postdocs who develop research software.
Fourth, we asked people reached by the above advertising to also forward the survey invitation within their own networks. 
Because of the solicitation approach we used, we cannot estimate how many people received our invitation.
We distributed the survey on August 27, 2019 and left it open for one month.

\begin{figure*}[!tb]
\input{fig_surveyquestions1}
\caption{Survey Questions Part-1}
\label{fig_survey_questions1}
\end{figure*}

\begin{figure*}[!tb]
\input{fig_surveyquestions2}
\caption{Survey Questions Part-2}
\label{fig_survey_questions2}
\end{figure*}

\subsection{Data Analysis}
\label{sec:Data-Analysis}
We anticipated the respondents might not have the background to answer all of the survey questions, therefore we allowed them to skip any question they did not want to answer.
Because we did not require respondents to answer all questions, we received some partially complete responses.
To ensure that we included only valid responses in our analysis, the first author manually went through each response to check its level of completeness.
We included responses that answered all quantitative questions and provided useful information in response to one or more qualitative questions.

We received both quantitative and qualitative data from the survey\footnote{The survey data is available in a public repository but set to private until publication of this paper~\citep{carver_eisty_2021}}.
We used the tool SPSS to analyze the quantitative data.
For any data that are qualitative, we used a grounded theory approach for the analysis. 
We individually coded the text using NVivo. 
Then we compared our results and identified the discrepancies.
We discussed all the discrepancies on an in-person meeting to solve any disagreements.
After that, we identified the relationships between the coded data and categorize them.
Finally, we categorized the categories into high-level core categories.
We used the programming language R to visualize the results into charts.

\section{Results}
\label{sec:Results}
This section describes the results from the 120 valid responses, as defined in Section~\ref{sec:Data-Analysis}.

\subsection{Demographics}
The following subsections detail each of the demographics gathered in Q1-Q5.

\subsubsection{Projects}
Q1 asked participants to optionally identify their research software project.
Respondents mentioned 66 unique research software projects.
Because of their project's privacy policies, 56 respondents did not name their project.
All but two project names were unique, with one identified by three respondents and the other by two.
To get a sense of project domains, we examined the website for each named project. 
The domains included: Physics, Chemistry, Biology, Geology, Astronomy, Mathematics, Climate Science, Neuroscience, Atmospheric science, Astrophysics, Computing infrastructure, Numerical Libraries, and Simulation tools.
Given the small number of respondents from any one domain, we were not able to analyze the effect of project domain on the results.
This result indicates that the respondents came from a wide variety of research software projects, which allows our overall findings to be useful to a broad audience of research software developers.

\subsubsection{Project Role}
\begin{figure}[!htb]
	\includegraphics[width=\textwidth]{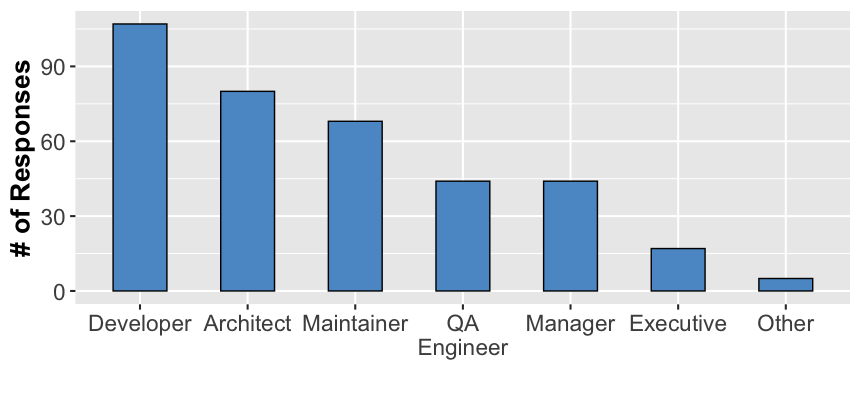}
	\caption{Respondents' role on project}
	\label{fig_role_on_project}
\end{figure}

Because people in different project roles likely have different perspectives on software quality and different experience with testing their projects, the respondent's role is an important demographic.
The results of Q2 (Figure~\ref{fig_role_on_project}) shows the distribution of respondents is skewed towards technical roles (e.g. Developer, Architect, and Maintainer).
Almost all respondents indicated Developer as at least one of their roles.
Note that because research software developers often hold multiple roles on a project, the survey allowed them to choose multiple responses.
Therefore, the sum of the bars in Figure~\ref{fig_role_on_project} is larger than the total number of respondents. 

\subsubsection{Project Experience}
\begin{figure}[!htb]
	\includegraphics[width=\textwidth]{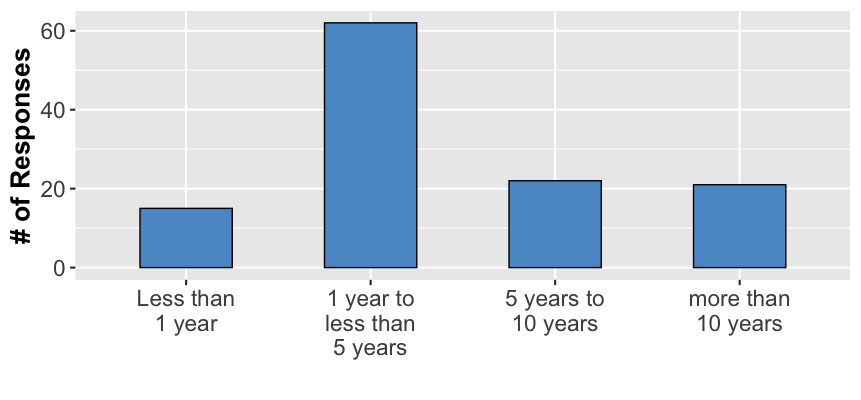}
	\caption{Number of years worked on research software}
	\label{fig_years_worked}
\end{figure}

A respondent's experience provides insight into whether she or he has enough knowledge to draw upon to provide helpful responses to the survey. 
The results from Q3 (Figure~\ref{fig_years_worked}) show the large majority of respondents had more than one year of experience in working research software projects.
Almost $1/3$ of had at least five years of experience, with $1/2$ of those having more than ten years of experience. 
This result shows that, the respondents had enough experience to provide valid responses to the survey.

\subsubsection{Project Stage}
\begin{figure}[!htb]
	\includegraphics[width=\textwidth]{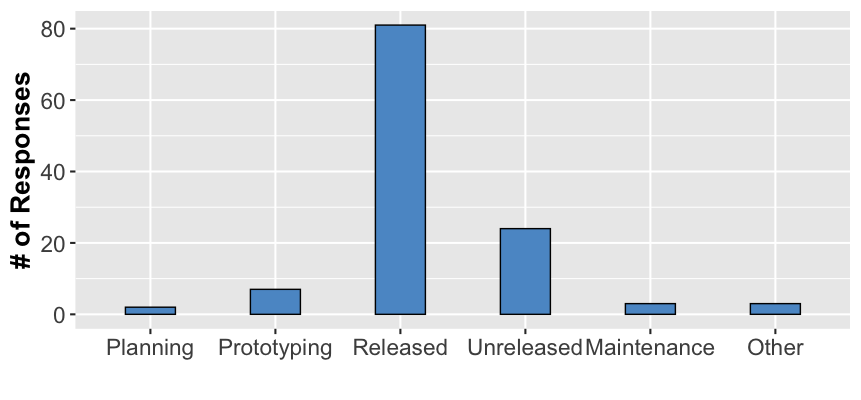}
	\caption{Project stage}
	\label{fig_development_stage}
\end{figure}

Because different types of testing are relevant at different stages of project development, the stage of the respondent's project could impact how he or she answers the survey questions.
The responses to Q4 (Figure~\ref{fig_development_stage}) show projects were overwhelmingly at the \textit{released} stage. 
This result is important because the projects at this stage should have already established testing practices in the project.

\subsubsection{Project Size}
\begin{figure}[!htb]
	\includegraphics[width=\textwidth]{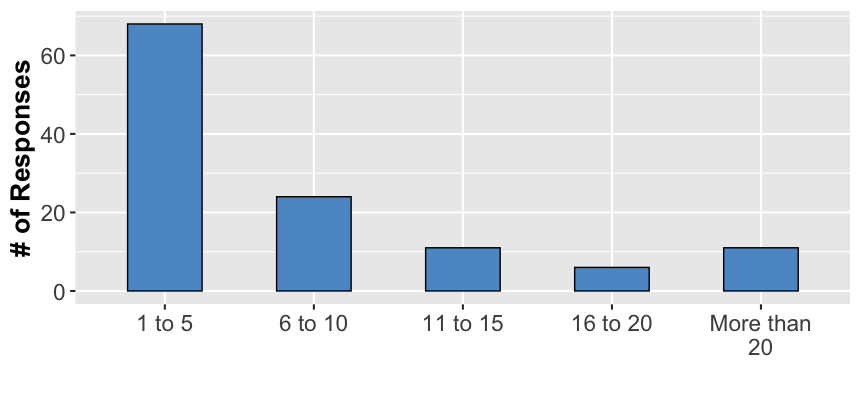}
	\caption{Number of developers}
	\label{fig_number_of_developers}
\end{figure}

Because teams of different sizes may have different perspectives on the use of software engineering practices~\citep{8588655}, we asked the respondents about their project size.  
The responses to Q5 (Figure~\ref{fig_number_of_developers}) shows, that while the largest group of respondents were on smaller teams, there is also a good distribution of respondents across larger team sizes.

\subsection{RQ1: What level of knowledge do research software developers have about software testing?}

\begin{figure}[!htb]
	\includegraphics[width=\textwidth]{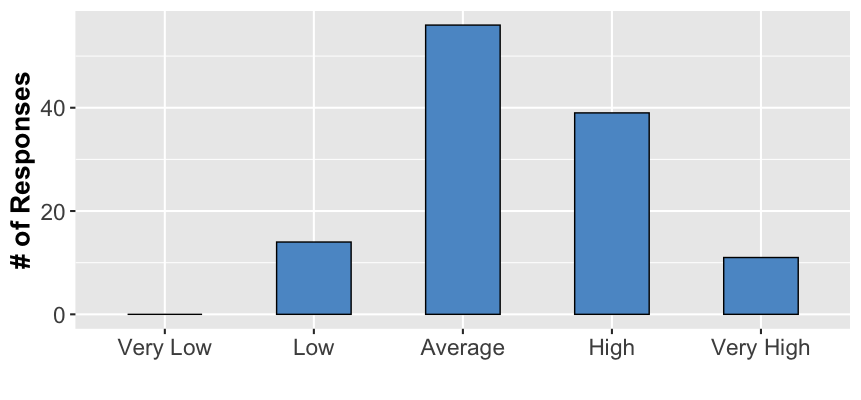}
	\caption{Confidence on knowledge of software testing}
	\label{fig_confident_on_knowledge}
\end{figure}

Survey question Q6 asked respondents how confident they were in their knowledge of software testing. 
The results in Figure~\ref{fig_confident_on_knowledge} show most respondents indicated they possessed at least an \textit{average} level of confidence about testing knowledge, with more than $1/3$ indicating \textit{high} or \textit{very high} confidence in their knowledge.
On a positive note, only a few respondents indicated they had \textit{low} confidence on knowledge, with none responding \textit{very low}.

\begin{figure}[!htb]
	\includegraphics[width=\textwidth]{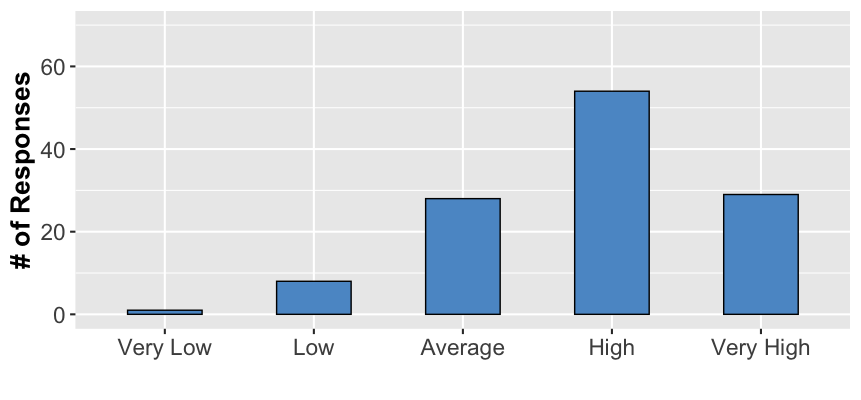}
	\caption{Level of understanding on the testing concepts used}
	\label{fig_understanding_concepts_used}
\end{figure}

\begin{figure}[!htb]
	\includegraphics[width=\textwidth]{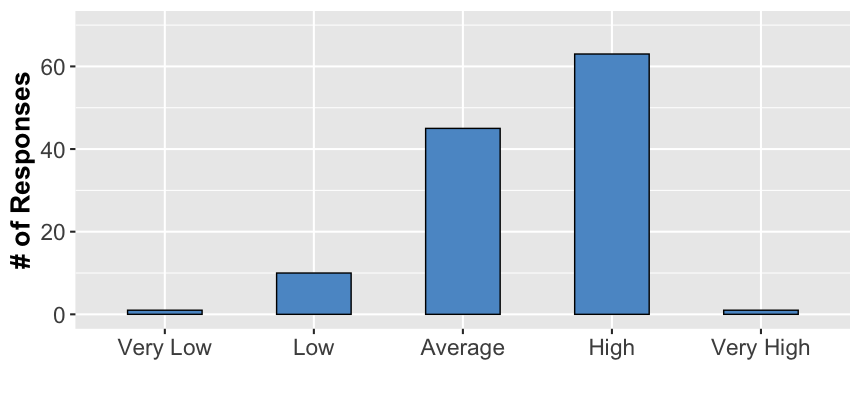}
	\caption{Level of understanding on the testing concepts needed}
	\label{fig_understanding_concepts_needed}
\end{figure}

Next, Q7 asked respondents about their level of understanding of the testing concepts they actually used in their projects.
The results in Figure~\ref{fig_understanding_concepts_used} show that more than half of the respondents indicated their level of understanding was \textit{high} or \textit{very high}.
Another $1/3$ indicated their understanding was \textit{average}.
Only a small number had a \textit{low} or \textit{very low} level of understanding.
Similarly, Q8 asked respondents about their level of understanding of the testing concepts needed for their projects, which might be different than those actually being used.
The results in Figure~\ref{fig_understanding_concepts_needed}, show a slightly different distribution than the previous question.
Most respondents still reported at least an \textit{average} level of understanding.
However, the responses shifted away from \textit{very high} into \textit{high} and \textit{average}.
These distributions are significantly different ($\chi^2$ = 31.0068, p $<$ .001).
Together, the responses to these two questions indicate that while survey respondents believed they had an adequate understanding of the testing concepts used in their projects, they were less confident about the testing concepts they actually needed.

To gain better insight into the respondents' level of knowledge, Q9 asked them to list any software testing techniques with which they were familiar.
By providing a free response question rather than a set of checkboxes, this question allowed us to judge the breadth of the respondents' testing knowledge.
Across all respondents, we collected a long list of testing techniques (see Appendix~\ref{sec:Appendix_2}).

\subsection{RQ2: How do research software developers test software?}

Given that the respondents had a reasonable understanding of testing (RQ1) and given the inherent difficulties in testing research software, the results for this question help us understand how testing occurs in practice.

\begin{figure}[!htb]
	\includegraphics[width=\textwidth]{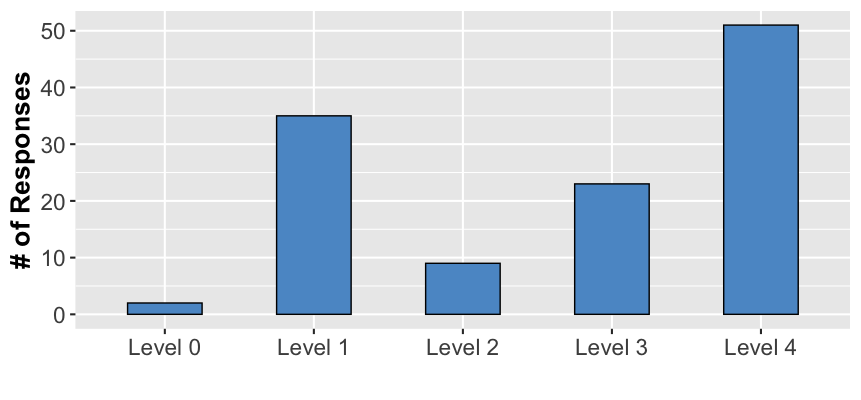}
	\caption{Goal of testing}
	\label{fig_goal_of_testing}
\end{figure}

Survey question Q10 asked respondents to choose which of the following testing goals (obtained from \textit{Introduction to Software Testing} by Ammann and Offutt~\citep{10.5555/3133461}) most closely matched their project: 
\begin{itemize}
    \item Level 0 - There is no difference between testing and debugging
    \item Level 1 - The purpose of testing is to show correctness
    \item Level 2 - The purpose of testing is to show that the software does not work
    \item Level 3 - The purpose of testing is not to prove anything specific, but to reduce the risk of using the software
    \item Level 4 - Testing is a mental discipline that helps all researchers develop higher quality software
\end{itemize}
The results (Figure~\ref{fig_goal_of_testing}) show the most common response is \textit{Level 4}, followed by \textit{Level 1}.
This result means respondents care about producing high-quality software to show correctness and doing so is a mental satisfaction to them.
It is encouraging because of having a concrete testing goal represents respondents' willingness to produce trustworthy software by employing proper testing on the projects.  

\begin{figure}[!htb]
	\includegraphics[width=\textwidth]{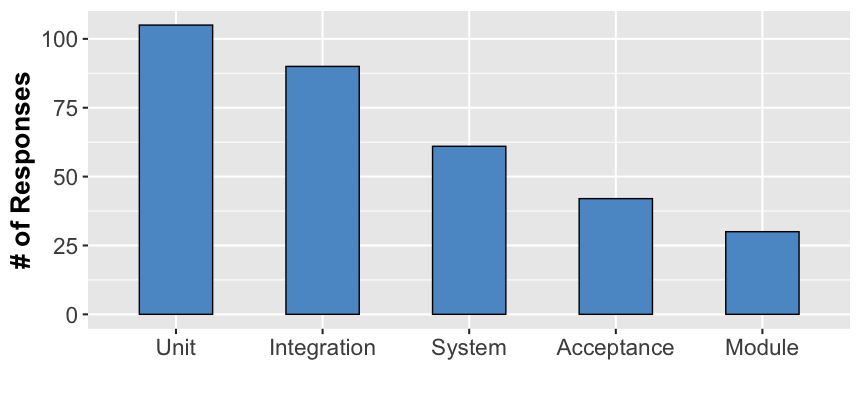}
	\caption{Testing methods used}
	\label{fig_testing_methods_used}
\end{figure}

\begin{figure}[!htb]
	\includegraphics[width=\textwidth]{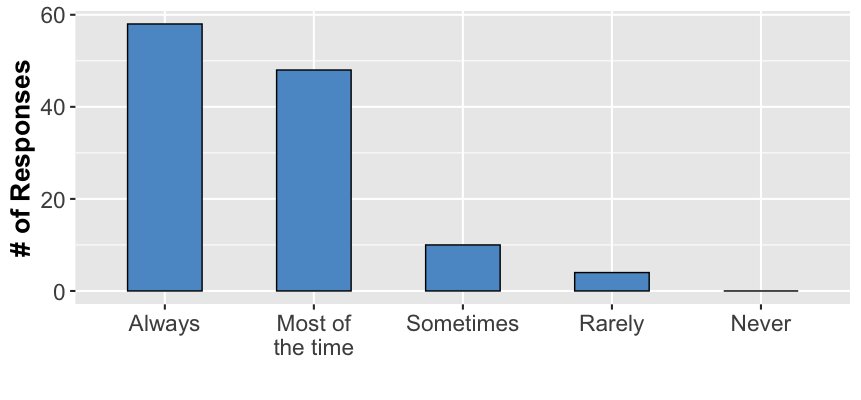}
	\caption{Usefulness of testing}
	\label{fig_usefulness_of_testing}
\end{figure}

The next question (Q11) asked respondents to indicate which testing methods (chosen from a list obtained from \textit{Introduction to Software Testing} by Ammann and Offutt~\citep{10.5555/3133461} and provided on the survey) their team uses.
The results in Figure~\ref{fig_testing_methods_used} show \textit{Unit Testing} is the most commonly used method among the respondents.
Many respondents also chose \textit{Integration Testing} and \textit{System Testing}.
The use of \textit{Acceptance Testing} and \textit{Module Testing} is less frequent.
Following on this question, Q12 asked respondents how useful they found software testing in their projects.
The results (Figure~\ref{fig_usefulness_of_testing}) show most found testing to be useful  \textit{always} or \textit{most of the time}.
Only a few respondents indicated testing was \textit{sometimes} or \textit{rarely} useful, with no one indicating it was \textit{never} useful.
These results indicate that respondents found some types of testing (system, unit, and integration) to be useful most of the time.

Last, Q13 provided an example list of testing techniques, along with their definitions, and asked respondents to indicate which of those techniques they actually use in their project.
In addition to the techniques provided on the survey, respondents could write in other techniques they use.
Table~\ref{tab:table-list_of_used_techniques} summarizes the responses.
The top portion of the table lists the testing techniques included on the survey.
The bottom portion of the table contains the techniques respondents provided in the \textit{other} section.

 \begin{table*}
\begin{center}
\caption{\label{tab:table-list_of_used_techniques}List of Used Testing Techniques.}
\begin{tabular}{ |p{.35\textwidth} | c || p{.3\textwidth}|c|  }
 \hline
 \multicolumn{4}{|c|}{Used Testing Techniques (from options)} \\
 \hline
 Name& Count &Name&Count\\
 \hline
 Assertion checking   & 94    & Error guessing   & 21\\
 Performance testing & 72  & Fuzzing test &  20\\
 Backward compatibility testing & 53 & Graph coverage &  17\\
 Statement coverage    & 53  & State transition & 17\\
 Test driven development&   50 & Logic coverage   & 11\\
 Condition coverage& 33  & Decision table based testing & 9\\
 Dual coding & 33    & Input space partitioning &   6\\
 Branch coverage   & 32     & Syntax-based   & 6\\
 Monte carlo test &   27 & Using machine learning &  4\\
 Boundary value analysis & 26 & Equivalence partitioning & 4\\
 Metamorphic testing &   26 & &\\
 \hline
 \multicolumn{4}{|c|}{Others (Write-ins) } \\
 \hline
 Regression testing   & 4    & Portability testing   & 1\\
 Bit-for-bit comparison & 1  & Code coverage &  1\\
 Benchmarking & 1 & Scaling test &  1\\ 
 \hline
\end{tabular}
 
\end{center}
\end{table*}

This result suggests that research software developers use a wide variety of testing techniques in their projects.
This result is also consistent with the results in Figure~\ref{fig_confident_on_knowledge} showing that respondents have an adequate level of knowledge and Figure~\ref{fig_understanding_concepts_used} showing they have good understanding of the testing concepts used in their projects. 

\subsection{RQ3: Why is testing research software difficult?}

\begin{figure}[!htb]
	\includegraphics[width=\textwidth]{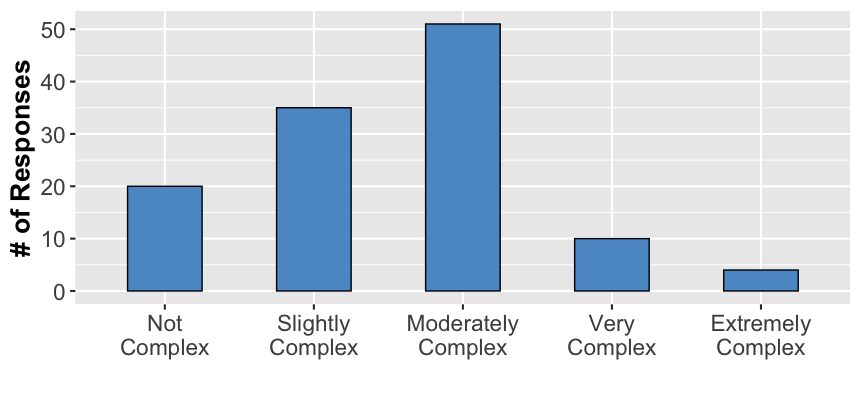}
	\caption{Complexity of testing research software}
	\label{fig_complex_to_test}
\end{figure}

To make any progress in overcoming the difficulties with testing research software, it is important to understand the specific reasons why testing is difficult.
This research question helps identify specific challenges and barriers.
Survey question Q14 asked respondents to rate the complexity of testing their projects.
According to the results (Figure~\ref{fig_complex_to_test}), the distribution of responses is skewed towards the lower complexity end of the scale, with the peak at \textit{Moderately Complex}.
This result means that while there is some level of complexity in testing research software, most respondents do not find the complexity to be too high.

\begin{figure}[!htb]
	\includegraphics[width=\textwidth]{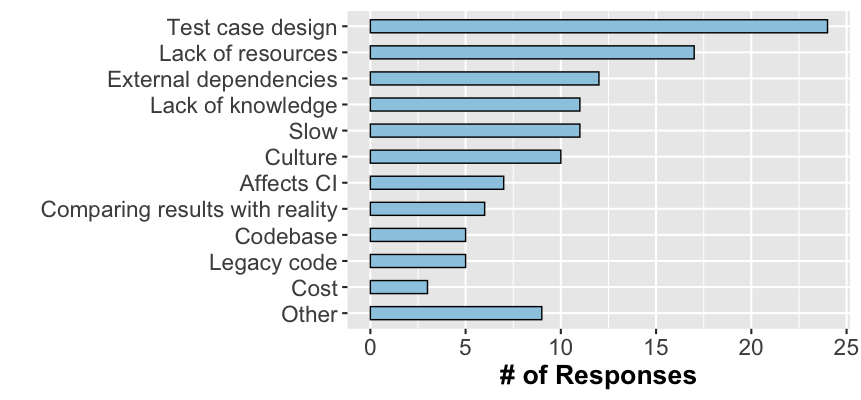}
	\caption{Challenges in testing research software}
	\label{fig_challenged_faced}
\end{figure}

To gain a deeper understanding of the difficulties with testing research software, Q15 asked respondents to explain any barriers or challenges they face with testing their software.
This question was open-ended.
Our qualitative analysis of these free-response answers resulted in the 12 high-level categories of challenges shown in Figure~\ref{fig_challenged_faced}.
The following text goes through each high-level challenge to explain what it means.

The most commonly mentioned class of challenges is \textbf{test case design}.
One respondent described the challenge as having difficulty \textit{``[e]ngineering good test cases and making sure that all equivalence classes of test cases are covered''}.
Another respondent indicated that \textit{``[a]pplication cases are usually too big/expensive to test, so breaking them down for meaningful system tests is a challenge.''}

The second most commonly mentioned challenge is the \textbf{lack of resources}, which includes, as one respondent said, \textit{``[l]ack of funding, people, and calendar time''}.
Another respondent commented that \textit{``[t]he amount of time needed to write tests gives them a bad reputation with developers who aren't convinced they are necessary''}.
Another respondent described the problem as \textit{``[l]imited amount of time is "allowed"/"allocated" for writing tests and setting up testing environments.''}

The third most commonly mentioned challenge is \textbf{external dependencies}.
The following is a representative response: 
\textit{``Tests should execute external proprietary software which is unavailable on services like TravisCI.''}

The fourth most commonly mentioned class of challenges is \textbf{lack of knowledge}.
One respondent described the problem as \textit{``..lack of testing knowledge -- collectively, as a team, we've probably heard of all of the possible ways of testing mentioned in the questions above, but several of these categories are not well understood by the team (or any individual within.''}

The fifth most commonly mentioned class of challenges is \textbf{slow}.
One respondent described this problem as \textit{``[t]ests take a long time to run, slows down continuous integration.''}

The sixth most commonly mentioned class of challenges is \textbf{culture}.
One respondent explained the challenge as \textit{``[m]ostly cultural, convincing my team mates that this is important to the sustainability of the code base''}.
Moreover, research software has \textit{``[a] culture that doesn’t value test development.''}

The seventh most commonly mentioned class of challenges is that testing \textbf{affects continuous integration}.
As a summary of these responses, one respondent indicated \textit{``[t]esting across multiple machines regularly is a challenge, due to the continuous integration tests running on a single machine.''}

The eighth most commonly mentioned class of challenges comes from the fact that \textbf{comparing results with reality} makes testing very difficult.
As a specific example, one respondent said \textit{``[s]ince we're developing a computational fluid dynamics code (it's an ocean model), the most difficult part was testing that the model produces the physically correct output.''}

The ninth most commonly mentioned challenge is a result of the \textbf{codebase} itself. 
One respondent describes the problem as \textit{``[t]he code has not been designed in a very modular way, so unit testing is not easy to implement.''}

The tenth most commonly mentioned challenge is \textbf{legacy code}.
Sometimes there are \textit{``[l]arge amounts of legacy code that were not developed with testing in mind.''}

The last class of challenges is \textbf{cost}.
Testing research software is difficult because of the \textit{``high cost in maintaining tests, expensive/slow to run full test suite.''} 

In addition to these high-level classes, there were a number of \textbf{other} challenges, including environmental changes, testing graphics production, and challenges related to programming languages and database systems. 

\subsection{RQ4: Is it possible to adapt existing testing methods to support the testing of research software?}

In this case, ``existing testing methods'' refer to the testing methods currently used in Commercial/IT software development.
These methods include unit, integration, system, acceptance, and module testing.
Gaining a better understanding of how these methods apply to the testing of research software will help research software developers have more confidence in the methods they can use and reduce the need to discover this information on their own.

\begin{figure}[!htb]
	\includegraphics[width=\textwidth]{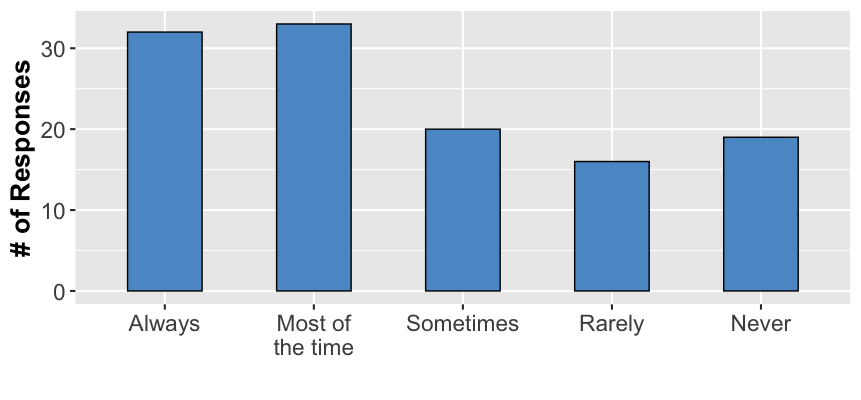}
	\caption{Applying Commercial/IT testing methods by team}
	\label{fig_apply_comm_it_methods_team}
\end{figure}

\begin{figure}[!htb]
	\includegraphics[width=\textwidth]{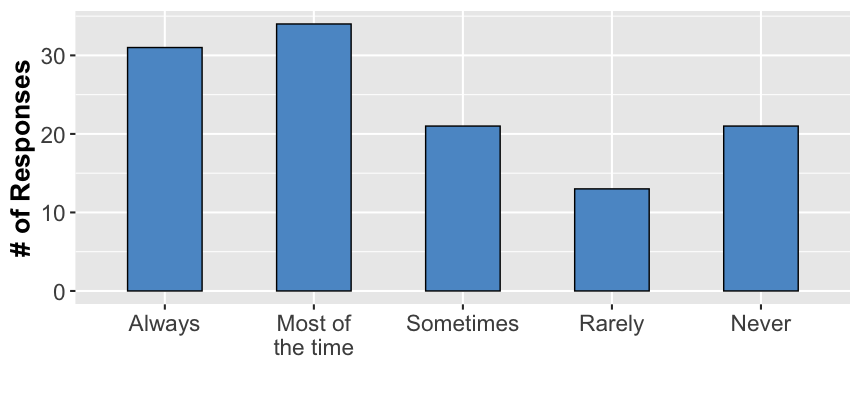}
	\caption{Applying Commercial/IT testing methods personally}
	\label{fig_apply_comm_it_methods_personally}
\end{figure}

\begin{figure}[!htb]
	\includegraphics[width=\textwidth]{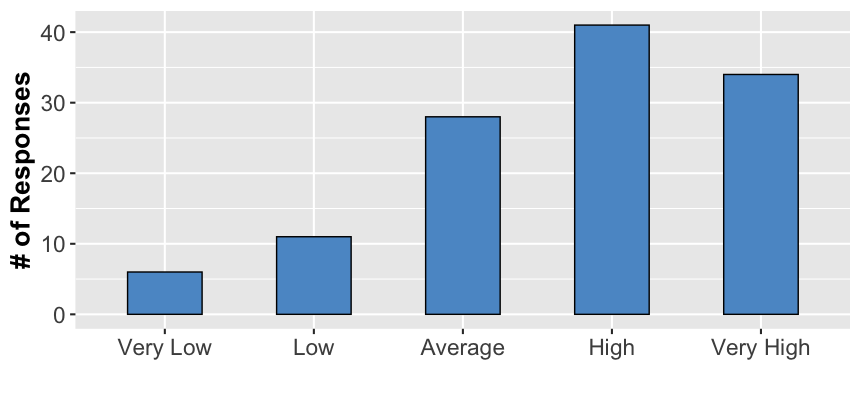}
	\caption{Value seen in using Commercial/IT testing methods}
	\label{fig_value_seen_comm_it_methods}
\end{figure}

We began with two survey questions about the frequency with which respondents use Commercial/IT testing methods as a team (Q16) and individually (Q17).
As Figures~\ref{fig_apply_comm_it_methods_team} and~\ref{fig_apply_comm_it_methods_personally} show most teams and individuals apply these methods at least \textit{sometimes}.
There is no significant difference between the distribution of responses.
Beyond whether the respondents use the Commercial/IT testing methods, Q18 ask the level of value they see personally in using these methods.
The distribution of results in Figure~\ref{fig_value_seen_comm_it_methods} show that the respondents generally saw value in using such methods, with more than half answering \textit{high} or \textit{very high}. 

\begin{figure}[!htb]
	\includegraphics[width=\textwidth]{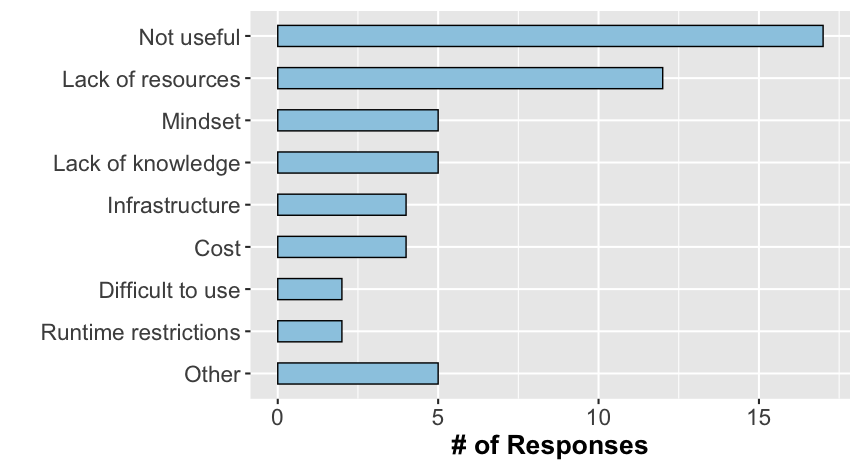}
	\caption{Challenges to adapt Commercial/IT testing methods}
	\label{fig_challenges_adapt_comm_it_testing}
\end{figure}

To gain insight into where the Commercial/IT methods cause problems, Q19 asked respondents to explain any challenges they faced in an open-ended manner.
Our qualitative analysis of the results identified the nine high-level categories of challenges shown in Figure~\ref{fig_challenges_adapt_comm_it_testing}.
Because we discuss some of these challenges in detail in response to other questions, we only highlight a subset here.

Overwhelmingly, the most common challenge respondents reported was that the methods were \textbf{not useful}.
This challenge arises because \textit{``[r]esearch software is typically not production software''} and \textit{``[s]ome commercial tools do not account for issues with numerical tolerances. Legacy codes are hard to get under test''}.
Another respondent indicated it is \textit{``Difficult to adapt [Commercial/IT testing methods] to the development of scientific software because of numerical errors and often not knowing the expected output''}.

The second most common challenge was \textbf{lack of resources}.
One respondent explained the challenge as \textit{``[l]ack of expertise, schedule demands, lack of R\&D (i.e. exploratory) s/w development oriented tools.''}

The third most common response is research software developer \textbf{mindset}.
One respondent summarized the problem as \textit{``[d]evelopers often feel like any time not spent developing code for the core software is not real work, thus time spent writing tests is less enjoyable''}.
In addition, the problem is \textit{``[c]ultural -- convincing people that it is beneficial, necessary, and worth their time and not insulting to their work''} 
There are also challenges that originate outside the team, such as \textit{``[o]ur funding agencies are generally not aware of their importance, and effort spent on testing is effort not spent writing/publishing papers''}.
Finally, there is a \textit{``[l]ack of familiarity with the ensemble of testing patterns and goals available.''}

\begin{figure}[!htb]
	\includegraphics[width=\textwidth]{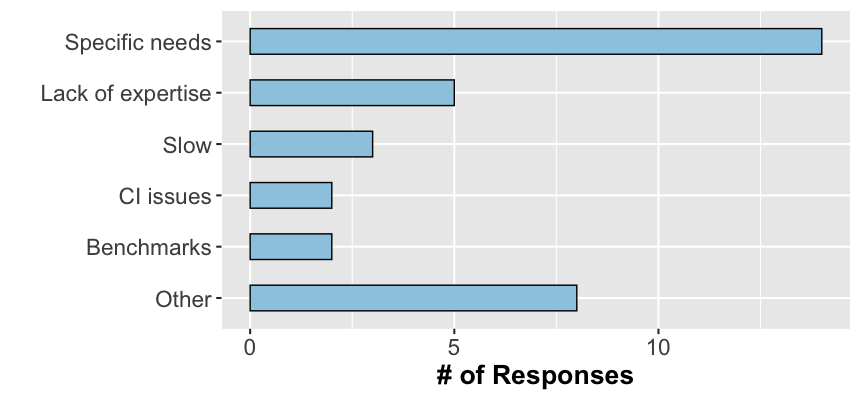}
	\caption{Challenges could not met by Commercial/IT testing methods}
	\label{fig_challenges_not_met}
\end{figure}

In addition to challenges with adapting Commercial/IT methods, there are cases where those methods are just not applicable. 
Survey question Q20 asked respondents to explain any challenges that could not be met by Commercial/IT testing methods.
The qualitative analysis produced five high-level challenges (Figure~\ref{fig_challenges_not_met}).

The most common challenge, by a large margin, was that the methods do not meet the \textbf{specific needs} of research software. 
These specific needs can include \textit{``[v]isualization [, i]mage processing \& analysis [, f]luid flow simulation [, s]ituations where there is no analytic solution or known correct answer [, i].e. no oracle''}.
A unique challenge of research software is that \textit{``[v]alidation requires domain expertise which is sometimes difficult to express in the commercial methods''}.
Another situation that is common in research software is whether the results are meaningful, as a respondent stated \textit{``IT methods are good for preventing errors / seg faults, but not good at catching if numbers are no longer meaningful (eg demand curves have inverted) or triaging to determine.''}

Similar to the responses to  other questions, the respondents also mentioned \textbf{lack of expertise}, \textbf{slow} to execute the test and \textbf{continuous integration issues} in response to this question. 
Finally, \textbf{benchmarks} is another challenge that could not be met by Commercial/IT testing methods.
One respondent indicated \textit{``[b]enchmarking scientific codes for accuracy (e.g. expected error convergence rates) is something that is not usually discussed in Commercial IT testing.''}

\subsection{RQ5: What improvements to the testing process do research software developers need?}

\begin{figure}[!htb]
	\includegraphics[width=\textwidth]{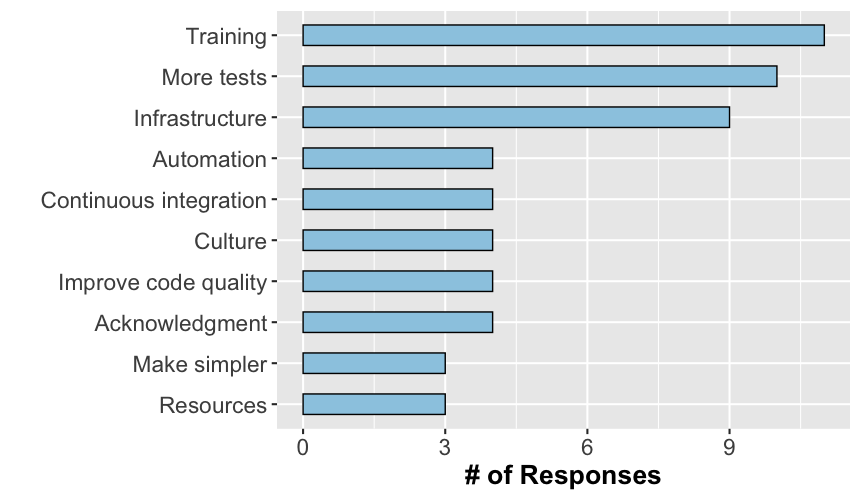}
	\caption{Improvement of the testing methods}
	\label{fig_improvement}
\end{figure}

Finally, to help provide guidance to the research software community, Q21 asked respondents to describe how to improve the testing process.
We consider the testing process in general here in this section. 
Then, in a follow-up study, we consider specific scenarios like test case design, test execution, testing tools, test coverage evaluation, and test quality.
The qualitative analysis produced ten high-level categories of improvements (Figure~\ref{fig_improvement}).

The most commonly mentioned improvement was proper \textbf{training}.
Some of the specific suggestions provided by respondents include: \textit{``[t]each it to grad students as an essential part of writing software''}, \textit{``[m]ore training, earlier on, in scientists' careers''}, \textit{``[s]eminars for practitioners, classes on scientific software engineering''}, and \textit{``[e]ducate scientists about the benefits of testing software''}.

Then the second most commonly mentioned improvement was \textbf{more tests}.
This improvement focuses on developer behavior to \textit{``[w]rite more unit tests, preferably using a dedicated testing framework as pFunit''} and through \textit{``[i]ncreased use of customized fuzz testing covering more input features.''}

The third most commonly mentioned improvement was \textbf{infrastructure}.
One respondent requested \textit{``[a] public service for testing that (1) is freely available for open source, (2) with many-tier (detailed, incremental) pricing structure for more machine time if needed, with (3) a sophisticated testing dashboard, similar to that of TeamCity''}.
Another respondents described the need as \textit{``[o]ur group needs more developers.  This is highest priority.  But, just as high:  we need support from the academic infrastructure providers to be able to run tests before our users hit the resources''}.
Another respondent wanted a \textit{``[simplified] infrastructure for creating new tests.''}

The fourth most commonly mentioned improvement was \textbf{automation}.
Specifically respondents wanted automation for \textit{``setting tests and analysis of results''} and \textit{``simpler methods for enabling/using tools.''}

Tied for the fourth most commonly mentioned improvement was \textbf{continuous integration}.
These systems need improvement because \textit{``[o]ur CI / testing system is very frail, and tends to break when there are system updates, requiring supervision and triage, and then there isn't a easy way to rerun tests on PRs that came in during the down time.''}

Also tied for the fourth most commonly mentioned improvement was changing the \textbf{culture} of testing in the research software developer community.
\textit{``Changing the culture.  Ensuring any pull requests that come in have adequate code coverage.  Incentivizing test development efforts.  Educating the developers on the importance of software testing and how best to go about it.  Making sure when bugs are getting fixed that the fix isn’t accepted until there’s a new unit test in place to cover the bug''}.
Similarily, there is a need to \textit{``[c]onvince other developers that are a part of the same project to adapt to these practices.''}

Also tied for the fourth most commonly mentioned improvement was the need to \textbf{improve code quality}. 
This improvement has to happen among the developers themselves, as one response suggested to \textit{``[h]ave HPC researchers, scientists, and engineers write better code. While improvements can certainly be made in testing, the vast majority of the problem currently is the state of the code written. You can't really test a 1000-line main() - you need structure, seams, and modularity in whatever language or paradigm you use''}.
Another respondent suggested that developers need to \textit{``Simplify[] code, reduce redundant tests, design better input variables, study other individual components and other possible problems with different architectures or compilers, etc.''}

Also tied for the fourth most commonly mentioned improvement is the need for proper \textbf{acknowledgement} to help motivate the research software developers to improve their testing.
There is a need for \textit{``[f]unding agencies and application users understanding the benefits of testing and support it enthusiastically''} and a need for \textit{``better incentive for others to contribute tests, period.''}

Moreover, initiatives to \textbf{make simpler} and provide adequate \textbf{resources} could potentially improve the testing process in research software development.

\section{Discussion}
\label{sec:Discussion}
This section discusses key insights from the detailed results relative to each of the overall research questions.

\subsection{RQ1 - Knowledge}
In general, research software developers are confident about their knowledge of software testing.
The survey respondents think they have at least an average level of understanding of the testing concepts used and needed in their projects. 
Our findings are somewhat inconsistent with the previous literature (Section~\ref{sec:research_questions}) which found that research software developers have little or no knowledge of software engineering. 
While it is possible that respondents overestimated their knowledge of software testing, there may also be a growing awareness of software engineering among the research software community.
This result is encouraging because as research software developers become more knowledgeable about software engineering practices, they will tend to produce higher-quality software. 
We need further study to better understand the source of the differences between our survey and the prior work in this area.

\subsection{RQ2 - Practices}
Research software developers have a clear goal for testing.
Many of the respondents viewed testing as a mental discipline that makes them confident in producing trustworthy results.
Again, this result is somewhat inconsistent with prior research (Section~\ref{sec:research_questions}) indicating that research software developers have no written quality goal.
Conversely, another set of respondents viewed the goal of testing as showing the correctness of the software, which evidences a much lower maturity level.
These inconsistent results show that there is still a large disparity in knowledge among the research software developer community.

The literature did not provide much evidence that research software developers use different testing methods~\citep{Kanewala:2014:TSS:2658281.2658307, Heaton-Carver:2015}.
However, the results in Figure~\ref{fig_testing_methods_used} show that research software developers do, in fact, use a wide variety of testing methods.
The respondents also reported using a wide variety of software testing techniques.
While the literature~\citep{Heaton-Carver:2015} suggests limited effectiveness of the testing practices currently used by research software developers, most of the respondents indicated testing was useful at least most of the time.
The increased use of testing could result from recent initiatives on applying software engineering practices to research software and an increased emphasis on making research software sustainable and reusable. 

\subsection{RQ3 - Difficulties}
In terms of challenges and barriers to testing research software, our results are consistent with the literature~\citep{Kanewala:2014:TSS:2658281.2658307}. 
The respondents indicated that there is some level of complexity in testing their projects.
The survey also identified a number of challenges for testing research software.
Test case design, lack of resources, and external dependencies are the most common challenges.
Lack of knowledge, test execution slowing down the development process, the lack of a proper testing culture all make testing difficult.  
Even though our results are consistent with the literature, we were able to contribute a list of concrete and current challenges taken directly from the experiences of research software developers. 

\subsection{RQ4 - Adapting existing testing methods}

In terms of applying existing Commercial/IT testing methods for testing research software projects, our results provide some insight where previous literature is deficient.
A previous literature review~\citep{Kanewala:2014:TSS:2658281.2658307} found very few papers that reported the use of any type of testing methods. 
Conversely, most of the respondents to our survey indicate that they apply Commercial/IT testing methods at least sometimes both personally and as a team.
They see at least an average level of value in applying Commercial/IT methods, while more than half of the respondents see high or very high value. 
While this result contradicts the published literature, it is encouraging. 
Applying available testing methods can be beneficial to many research software developers. 

Even with the positive view of Commercial/IT testing techniques, there are still some challenges to adapting those techniques for use in research software.
In many cases, Commercial/IT testing techniques are often not useful.
Respondents identified specific needs of research software that could not be met by the existing testing methods.
In spite of the challenges to adapting these testing methods for use in research software, this result is encouraging because it indicates that many research software developers are trying to apply these methods.

\vspace{-8pt}

\subsection{RQ5 - Improvement}
We identified many potential improvements to the testing process for research software. 
These results can serve as guidance to software teams who want to incorporate good practices as well as to testing researchers who want to provide more appropriate techniques for research software.

One critical need is formal training on fundamental testing concepts and the importance of testing. 
The training should provide research software developers with hands-on experience so they are able to understand and utilize existing testing techniques and tools in their projects, where appropriate.
As research software developers better understand the existing tools and techniques, they will also be able to identify gaps that can be filled by modifying existing tools and techniques or by creating new ones.

There is a need for more tests, better infrastructure, and test automation.
Incorporating more tests increases the changes of having correct results.
Developers of research software may limit their testing because of lack of infrastructure to ease the process. 
Proper infrastructure support and test automation can help motivate research software developers to test their projects well.
Moreover, changing the culture of research software to one that embraces testing, improved code quality, and proper acknowledgment for testing effort will help improve the overall testing process.


\section{Threats}
\label{sec:Threats}
This section describes the threats to validity of the study.

\subsection{Internal Threats}
The primary threat to internal validity is whether participants understood the software engineering concepts in the same way we intended them.
If the respondents did not understand the concepts or had different definitions, then the results would be less reliable.
Because the members of the target survey population are not traditional software engineers, it is possible that they lacked the necessary knowledge to properly answer the questions.

We provided appropriate definitions of terms to minimize confusions of the respondents.
Because we did not observe any misunderstandings and inconsistencies in the free-response questions, we find this threat is minimal.

\vspace{-16pt}

\subsection{External Threats}
If the survey respondents are not representative of the population of research software developers, the results are less generalizable.
In addition, it is not possible to measure the size of a community as diverse as the research software community.
So, it is possible that the sample included in our study is not representative of the overall population.
To reduce this threat, we recruited participants from different countries and projects. 
While it is clear that all participants are research software developers, some of their responses suggest they may be more interested in software testing than the average research software developer.
In addition, they took time to answer a survey about testing.
Therefore, the responses may be biased towards developers who are already predisposed towards the use of testing.
In addition, respondents' self-assessment may be wrong. 
They may have overestimated or underestimated their knowledge relative to the survey questions. 

\subsection{Construct Threats}
The primary construct validity threat is the participants may have misunderstood the questions. 
We took great care in writing the survey questions and verified them by expert research software developers and software engineering researchers.
In addition, we provided enough definitions to the respondents without biasing them so they could use their own judgment to respond. 

\subsection{Conclusion Threats}
It is possible that, with additional information, other conclusions could be drawn from the data gathered in this study.
We rely on the participants' perceptions of software testing, which may not match reality.
Another potential threat is that we used a standard software engineering textbook~\cite{10.5555/3133461} to define the testing terminology in the survey.
There are other sources, including SWEBOK~\cite{SWEBOK2014} or ISTQB\footnote{\url{https://www.istqb.org/}}, which have different perspectives.
While the textbook and these other sources are not mutually exclusive, there is a chance that survey participants could have different understandings of the terminology.

\section{Conclusion and Future work}
\label{sec:Conclusion}
Testing is an essential practice for building high-quality and trustworthy software. 
Because research software often supports critical situations and produces evidence for research publications, it is important for researchers to use appropriate testing approaches to complement their development methods. 
To gain insight into the practice of testing for research software, we conducted a survey of practicing research software developers.
The results from this survey help other research software developers understand how their peers test their software.

This paper report insights about testing research software gained from 120 responses to a survey of research software developers. 
The paper discusses the overall knowledge of software testing among research software developers, current practices of testing, difficulties in testing research software, challenges in adapting  existing testing methods, and potential improvements to the testing process.

Research software developers are somewhat confident (Figure~\ref{fig_confident_on_knowledge}) in their knowledge of testing and report a level of average or higher (Figure~\ref{fig_understanding_concepts_used}) for their understanding of the testing concepts used and needed in their projects.
Research software developers have clear testing goals and find  many types of testing techniques useful.
However, respondents also report a number of challenges and barriers for testing research software, including: 
test case design, lack of resources, external dependencies, and lack of knowledge.
In addition, it is not always easy for research software developers to use of Commercial/IT testing techniques.
Providing proper training and creating a culture that values testing could address many of these difficulties and improve the testing process. 


Given the fact that our results were drawn from a convenience sample and the fact that there is much diversity in the research software space, there is a need for further study to verify these findings in other contexts.
In our future work, we will study the specific technical challenges of testing research software.
We will analyze how testing practices and methods differ across project types and domains.
Once we identify the specific technical challenges, we will better understand how to develop new testing approaches for research software. 
Gathering this type of information will require close interaction with research software teams across different domains.
Ultimately, this work will allow us to develop best practices around testing that will be of value to the research software community.

\begin{acknowledgements}
We thank the study participants and NSF-1445344.
\end{acknowledgements}

\appendix
\section{Definitions Provided}
\label{sec:Appendix}
We refer to Figures~\ref{fig_survey_questions1} and~\ref{fig_survey_questions2} for the survey question.
In this section we listed the definitions we provided in the actual survey.

\begin{itemize}
    \item \textbf{Acceptance testing} - Assess software with respect to requirements or users’ needs.
    \item \textbf{Architect} - An individual who is a software development expert who makes high-level design choices and dictates technical standards, including software coding standards, tools, and platforms.
    \item \textbf{Assertion checking} - Testing some necessary property of the program under test using a boolean expression or a constraint to verify.
    \item \textbf{Backward compatibility testing} - Testing whether the newly updated software works well with an older version of the environment or not.
    \item \textbf{Branch coverage} - Testing code coverage by making sure all branches in the program source code are tested at least once.
    \item \textbf{Boundary value analysis} - Testing the output by checking if defects exist at boundary values.
    \item \textbf{Condition coverage} - Testing code coverage by making sure all conditions in the program source code are tested at least once.
    \item \textbf{Decision table based testing} - Testing the output by dealing with different combinations of inputs which produce different results.
    \item \textbf{Developer} - An individual who writes, debugs, and executes the source code of a software application.
    \item \textbf{Dual coding} - Testing the models created using two different algorithms while using the same or most common set of features.
    \item \textbf{Equivalence partitioning} - Testing a set of the group by picking a few values or numbers to understood that all values from that group generate the same output.
    \item \textbf{Error Guessing} - Testing the output where the test analyst uses his / her experience to guess the problematic areas of the application.
    \item \textbf{Executive} - An individual who establishes and directs the strategic long term goals, policies, and procedures for an organization's software development program. 
    \item \textbf{Fuzzing test} - Testing the software for failures or error messages that are presented due to unexpected or random inputs.
    \item \textbf{Graph coverage} - Testing code coverage by mapping executable statements and branches to a control flow graph and cover the graph in some way.
    \item \textbf{Input space partitioning} - Testing the output by dividing the input space according to logical partitioning and choosing elements from the input space of the software being tested.
    \item \textbf{Integration testing} - Asses software with respect to subsystem design.
    \item \textbf{Logic coverage} - Testing both semantic and syntactic meaning of how a logical expression is formulated.
    
    \item \textbf{Maintainer} - An individual who builds source code into a binary package for distribution, commit patches or organize code in a source repository.
    \item \textbf{Manager} - An individual who is responsible for overseeing and coordinating the people, resources, and processes required to deliver new software or upgrade existing products.
    \item \textbf{Metamorphic testing} - Testing how a particular change in input of the program would change the output.
    \item \textbf{Module testing} - Asses software with respect to detailed design.
    \item \textbf{Monte carlo test} - Testing numerical results using repeated random sampling.
    \item \textbf{Performance testing} - Testing some of the non-functional quality attributes of software like Stability, reliability, availability.
    \item \textbf{Quality Assurance Engineer} - An individual who tracks the development process, oversee production, testing each part to ensure it meets standards before moving to the next phase.
    \item \textbf{State transition} -  Testing the outputs by changes to the input conditions or changes to 'state' of the system.
    \item \textbf{Statement coverage} - Testing code coverage by making sure all statements in the program source code are tested at least once.
    \item \textbf{Syntax-based testing} - Testing the output using syntax to generate artifacts that are valid or invalid.
    \item \textbf{System testing} - Asses software with respect to architectural design and overall behavior.
    \item \textbf{Test driven development} - Testing the output by writing an (initially failing) automated test case that defines a desired improvement or new function, then produces the minimum amount of code to pass that test.
    \item \textbf{Unit testing} - Asses software with respect to implementation.
    \item \textbf{Using machine learning} - Testing the output values using different machine learning techniques.
    \item \textbf{Using statistical tests} - Testing the output values using different statistical tests.
\end{itemize}

\section{List of Testing Techniques}
\label{sec:Appendix_2}
This appendix provides the list of testing techniques respondents mentioned they were familiar with in response to the survey question Q9.
The numbers in the parenthesis represent how many respondents indicated that testing technique. 

\subsection{Testing Methods}
Acceptance testing (9),
Integration testing (43),
System testing (14),
Unit testing (87)

\subsection{Testing Techniques}
A/B testing (1),
Accuracy testing (1),
Alpha testing (1),
Approval testing (2),
Answer testing (1),
Assertions testing (3),
Behavioral testing (1),
Beta testing (1),
Bit-for-bit (1),
Black-box testing (4),
Built environment testing (1),
Builtd testing (1),
Checklist testing (1),
Checksum (1),
Compatibility Testing (1),
Concolic testing (1),
Correctness tests (1),
Dependencies testing (1),
Deployment testing (1),
Dynamic testing (3),
End-to-end testing (2),
Equivalence class (1),
Engineering tests (1),
Exploratory tests (1),
Functional testing (6),
Fuzz testing (12),
Golden master testing (1),
Install testing (1),
Jenkins automated testing (1),
Load testing (1),
Manual testing (2),
Memory testing (6),
Mock testing (6),
Mutation testing (5),
Penetration testing (1),
Performance testing (6),
Periodic testing (1),
Physics testing (1),
Property-based testing (2),
Random input testing (2),
Reference runs on test datasets (1),
Regression testing (39),
Reliability testing (1),
Resolution testing (1),
Scientific testing (2),
Security testing (1),
Smoke test (2),
Statistical testing (1),
Stress test (1),
Usability testing (1),
Use case test (1),
User testing (2),
Validation testing (9),
White-box testing (2)

\subsection{Testing Tools}
CTest (3),
gtest (1),
jUnit (1)

\subsection{Other types of QA}
Code coverage (16),
Code reviews (2),
Documentation checking (3),
Static analysis (6)

\subsection{Others}
Agile (1),
Asan (1),
Automatic test-case generation (1),
Behavior-Driven Development (1),
Bamboo (1),
Benchmarking (1),
Caliper (1),
Code style checking (1),
Coding standards (1),
Comparison with analytical solutions (1),
Continuous integration (33),
Contracts (1),
DBC (1),
Design by contract (1),
Doctests (1),
Formal Methods (2),
GitLab (1),
License compliance (1),
Linting (1),
Method of exact solutions (1),
Method of manufacture solution (2),
Monitoring production apps (1),
Msan (1),
N-version (1),
Nightly (1),
Pre-commit (1),
Profiling (1),
Release (1),
Run-time instrumentation and logging (1),
Squish (1),
Test-driven development (18),
Test suites (1),
Tsan (1),
Visual Studio (2)

\bibliographystyle{spbasic}
\bibliography{main}

\end{document}